\documentstyle[12pt,epsfig]{article}
\begin{document}
\begin{titlepage}
\hspace*{\fill}{IMSc-94/52}
\vspace*{\fill}
\begin{center}
{\Large \bf Infrared Behaviour of Systems With Goldstone Bosons}\\[1cm]
Ramesh Anishetty, Rahul Basu, N.D. Hari Dass and 
H.S.Sharatchandra\footnote{email:ramesha,rahul,dass,sharat@imsc.ernet.in}\\
{\em The Institute of Mathematical Sciences, Madras-600 113, INDIA}
\end{center}
\vspace{2cm}
\begin{abstract}
We develop various complementary concepts and techniques for 
handling quantum fluctuations of Goldstone bosons.We 
emphasise that one of the consequences of the masslessness of 
Goldstone bosons is that
the longitudinal fluctuations also have a diverging susceptibility
characterised by
an anomalous dimension  $(d-2)$ in space-time dimensions 
$2<d<4$.In $d=4$ these fluctuations diverge
logarithmically in the infrared region.We show the generality
of this phenomenon on the basis of 
i). Renormalization group flows, ii). Ward identities, and
iii). Schwinger-Dyson equations.We also obtain an explicit form for the 
generating functional of one-particle irreducible vertices
of the O(N) (non)--linear $\sigma$--models in the leading
$1/N$ approximation.We show that this incorporates all infrared
behaviour correctly both in linear and non-linear $\sigma$--
models. These techniques provide an alternative to chiral
perturbation theory.Some consequences are discussed briefly.
\end{abstract}
\vspace*{\fill}
\end{titlepage}
\section{Introduction}

The concept of spontaneous breaking of a continuous global symmetry
with the attendant Goldstone phenomenon of massless scalar 
excitations has deeply influenced many developments in field 
theory and statistical physics. The existence of massless modes
can already be demonstrated in Landau - Ginzberg mean field 
analysis (or tree level, in the field theory jargon).
However, it is not sufficient to stop at this level. Precisely 
because they are massless, the quantum/statistical fluctuations 
of the Goldstone bosons may significantly alter the conclusions
drawn from mean field theory.Proofs of the Goldstone theorem
\cite{gth}
in quantum field theory have traditionally been used only to 
establish the existence of massless
poles coupling to the currents of broken symmetry. They have not 
usually focussed on the correlation functions. Most of the
applications in quantum field theory have been based on the 
understanding at tree level. Notable exceptions are 
the soft pion theorems \cite{spt}
combining PCAC with current algebra.These are 
mostly applied to extract the leading infrared behaviour. 
A general analysis of these issues based essentially only on symmetry arguments
was given some time ago by Weinberg \cite{sw}. In that article he also
proposed methods to go beyond the leading order results. He also applied
renormalization group methods to understand some universal features of these
results. Quite recently, some of Weinberg's ideas have been extensively used
in the context of the so called ``chiral perturbation theory''\cite{cpt}.

In contrast, in statistical physics, powerful techniques 
have  been developed for handling the fluctuations
in this context. These  are of importance for quantum field 
theory also.Some examples are: i)the finite temperature chiral 
symmetry restoring transition in QCD, where Goldstone phenomenon
in three Eucilidean dimensions is of relevance \cite{wilc};
ii)QCD at temperatures below the chiral transition temperature as well as
zero temperature QCD, where Goldstone phenomenon in 4 euclidean dimensions
is of relevance,and
iii)the problems that are handled by chiral perturbation theory at 
present, i.e.  effects of pion loops on various processes
\cite{cpt}.

In this paper we point out various complimentary concepts 
and techniques for handling the effects due to quantum fluctuations 
of Goldstone bosons. We develop techniques for computing 
correlation functions away from the infrared regime.We 
show that there are 
certain universal features, valid even in four space-time 
dimensions.Our approach provides an alternate procedure for
chiral perturbation theory which is closer in its spirit to what has been
proposed by Weinberg on the basis of RG equations.
This aspect is elaborated upon in section 7.

\section{Infrared dynamics of pions and sigma}

Often the paradigm in quantum field theory is what is seen
from a tree level analysis of the O(N) linear and non-linear
$\sigma$-models.  The linear $\sigma$-model is given by the action
\cite{z},
\begin{equation}
S =  \int d^dx(\frac{1}{2} (\partial_\mu \vec\Phi(x))^2
   -\frac{U}{4} (\vec\Phi^2(x)-C^2)^2)
\end{equation}
$\vec \Phi=\{ \Phi_i,i=1,2,....N \}$ 
transforms as the fundamental representation of O(N).
If $C^2>0$, the minimum of the classical potential is 
at $\vec \Phi^2 =C^2$. Therefore the VEV $<\vec \Phi>$ is no longer zero
in a tree level analysis.Choosing the vacuum
$<\Phi_N>=C, <\Phi_i> =0, i=1,2,...(N-1)$,
among the equivalent vacua, we get,

\begin{eqnarray}
S &=& \int d^dx (\frac{1}{2} (\partial_{\mu} \vec \pi(x))^{2}
+\frac{1}{2} (\partial_{\mu} \sigma(x))^{2} -UC^2 \sigma^2(x)  \nonumber \\ 
& & -UC \sigma(x) (\sigma^2(x)+\vec \pi^2(x))
-\frac{U}{4} (\sigma^2(x)+\vec \pi^2(x))^2
\end{eqnarray}
where $\sigma= \Phi_N -<\Phi_N>,\vec \pi= \{\Phi_i, i=1,2,...(N-1) \}$.
Therefore at tree level, we have $(N-1)$ massless pions 
and a massive $\sigma$  with a mass $\sqrt {2UC^2}$.

Precisely because the Goldstone bosons are massless, loop
corrections may drastically alter the above picture.
Now it is well known that infrared divergences due to 
massless fluctuations drastically alter 
the conclusions of a semi-classical analysis, especially if the
spacetime dimension d is less than four \cite{z}.Already, 
way back in 1940, Holstein and Primakoff \cite{hp} argued that not only
the transverse mode (i.e. $\pi$) but also the longitudinal
mode ($\sigma$) is soft in a quantum ferromagnet in three dimensions. 
Their arguments can be understood,as follows, 
in terms of the non-linear $\sigma$ -model.

The non-linear $\sigma$- model is relevant to the infrared dynamics
of Goldstone bosons 
\cite{z} .A simple heuristic argument is that if $\sigma$ is massive, 
then for the study of the 
infrared dynamics of the pions, we may use an effective Lagrangian 
obtained by integrating  over the heavy $\sigma$ degree of freedom.
Also for leading infrared behaviour we may retain only the
lowest (two) derivative terms in the effective action.The result is 
simply the Heisenberg model,

\begin{equation}
S_{NL} = 
\frac{1}{2t} \int d^dx (\partial_{\mu} \vec \phi (x))^2
\end{equation}
with the constraint, $\vec \phi^2(x)=1$.
Statistical physics in d space dimensions is simply the
Euclidean quantum field theory in $d$ space-time dimensions.
In statistical physics $t$ is proportional to the temperature,
whereas in quantum field theory it is related to the pion decay
constant, $t=2f_{\pi}^{-2}$(in what follows, we set $f_\pi= 1$).
If the spins are ordered in the direction of $\phi_N$, then 
eliminating this field using the constraint, 
the action can be written entirely in terms of 
$\vec \pi= \{\phi_i, i=1,2,...(N-1) \}$. They describe
the spin waves or the transverse
fluctuations.At the tree level for the $\pi$ propagator
we get,

\begin{equation}
\Delta_{\pi} (k)=k^{-2}
\end{equation}

We may also consider the correlations for longitudinal 
fluctuations i.e. of the field
$\sigma= \phi_N  \sim :\sqrt {1-\vec \pi^2}:-1$(: refers to any 
ordering prescription that makes products of fields at the 
same point meaningful).
This means,
$\sigma \sim :\vec \pi^2:+\frac{1}{4}:\vec \pi^4:+.....$.
We may expect the leading 
infrared behaviour to come from the first term.
This would mean that so far as the leading infrared behaviour 
is concerned, \cite{wal}, \cite{lee}
 $\sigma \sim :\bf \pi^2:$.  If we ignore
self-interactions of $\bf \pi$, and use the above propagator for
$\bf \pi$ we get,for the $\sigma$ propagator,

\begin{equation}
\Delta_{\sigma} (x)= \Delta_{\pi}^2 (x)=
\mid {x} \mid ^{4-2d}
\end{equation}
This gives,  atleast for $d<4$,in momentum space,
\begin{equation}
\Delta_{\sigma} (k)= \mid {k} \mid ^{d-4}
\end{equation}
Ignoring the self interactions of $\pi$'s
will later be seen to be  justified because these are soft 
in the infrared region--
a consequence of the soft pion theorems.The result
is that in $d=3$ the $\sigma$--propagator is
$\mid {k} \mid ^{-1} $, to be contrasted with the 
$\pi$-propagator.This means that $\sigma$ also has
long range correlations.
This appears to negate the naive deduction of the  
non-linear $\sigma$ model obtained by integrating over the heavy 
$\sigma$ field in the linear $\sigma$ model \cite{wein}.

One could dismiss these results as artifacts of a 1-loop
calculation. In fact the infrared divergences grow with the 
loop order. Therefore it is imperative to tackle this issue
in a framework that transcends loop expansion.There is an
exactly solvable model, the Berlin-Kac model \cite{bk} which goes beyond 
loop-expansion and again gives precisely the same diverging 
longitudinal susceptibility and also a massless $\sigma$
in $d<4$. Again it is not clear whether this is an artifact
of the model.

In order to get a handle on the infrared dynamics of the $\sigma$ field,
Pokrovsky and Patashinky \cite{pp}  proposed a 
"Conserved Modulus Principle"
(CMP).This principle holds that the fluctuations are such as to 
maintain the modulus of the N- component order parameter. 
In mathematical terms CMP states that\\
\begin{equation}
2<\phi>  \delta \phi_L + \delta \phi_T^2 = 0 \label{pp1}
\end{equation}
Now the expectation value of $\delta \phi_T^2$ is given by the 
Ornstein- Zernike form:
\begin{equation}
\delta \phi_T^2 = \int {d^dq\over{q^2+H/<\phi>}} \label{pp2}
\end{equation}
It is convenient to use susceptibility to describe the infrared
behaviour.The longitudinal and the transverse susceptibility in
presence of an external magnetic field $\vec H =(0,0,...,0,H)$
are given by ,

\begin{equation}
\chi_L =lim_{k \rightarrow 0} \Delta_{\sigma} (k^2),\ \  
\chi_T =lim_{k \rightarrow 0} \Delta_{\pi} (k^2) 
\end{equation}
Using Eqns. \ref{pp1} and \ref{pp2}, we get, for $d<4$,

\begin{equation}
\chi_L = {\partial(\delta \phi_L)\over \partial H} 
\sim (H)^{d/2-2}~~~~~~\chi_T\sim H^{-1}
\end{equation}
Thus the longitudinal suceptibility also diverges.
Patashinsky and Pokrovski have offered a proof for CMP in \cite{ppb}. 
They express the change in the thermodynamic
potential U computed to quartic order in the fluctuations as\\
\begin{equation}
\delta U =1/2[{(\delta \phi_{T})^2\over \chi_{T}} + {(\delta
\vec\phi^2)^2\over 4 \chi_{L}\bar \phi^2}]
\end{equation}
and interpret this as implying CMP.
Though to quartic order the correct form of this variation,
which differs from the one given above (and which can be verified using
the Ward identities discussed later), is
\begin{equation}
\delta U ={1\over 2 \chi_{T}}[(\delta \phi_{T})^2+(\delta \phi_{L})^2-{(\delta \vec\phi
^2)^2\over 4 \bar \phi^2}] + {1\over 2 \chi_{L}}[{(\delta
\vec\phi^2)^2\over 4 \bar \phi^2}]
\end{equation}
the fact that CMP is equivalent to the correct statement that the
extreme IR behaviour is governed by the non-linear $\sigma$-model
indicates the possibility of a better proof of CMP.\\

The new era for understanding the role of fluctuations of Goldstone
bosons started with the renormalization group techniques.
Brezin,  Wallace and Wilson \cite{bww} computed the equation of state for O(N)
$\sigma$--model using the $\epsilon$--expansion.They found important 
differences from the corresponding calculation for the N=1 case
(i.e. the Ising model).This
could be traced to the infrared singularities due to the 
Goldstone bosons that are present everywhere in the coexistence 
region for $N>1$. In particular it meant that the conclusions of the 
tree level analysis could not be right (atleast for $d<4$).

At the critical point ($T=T_c, H=0$), $\sigma$ and $\pi$ are on 
the same footing. Both their  susceptibilities
diverge with  vanishing magnetic field as,

\begin{equation}
\chi_T= \chi_L \sim H^{{1\over\delta}-1}
\end{equation}

\noindent governed by the critical exponent $\delta$. The question 
of interest
is the behaviour for $T<T_c$ as $H \rightarrow 0$.The assumption 
that the singularities due to Goldstone bosons exponentiate in the 
$\epsilon$--expansion suggested that,

\begin{equation}
\chi_T \sim H^{-1}, \quad  \chi_L \sim H^{-\epsilon/2}
\end{equation}

Holstein -Primakoff \cite{hp}, Berlin-Kac \cite{bk}, and 
Patashinski-Pokrovski \cite{pp} models have all made the same 
prediction.  Such a result has also been proved rigorously for O(2) case
using correlation inequalities in Ref. \cite{new}.

A deeper understanding can be obtained by considering the
renormalization group flows shown in the figure below:
\begin{figure}[htb]
\begin{center}
\mbox{\epsfig{file=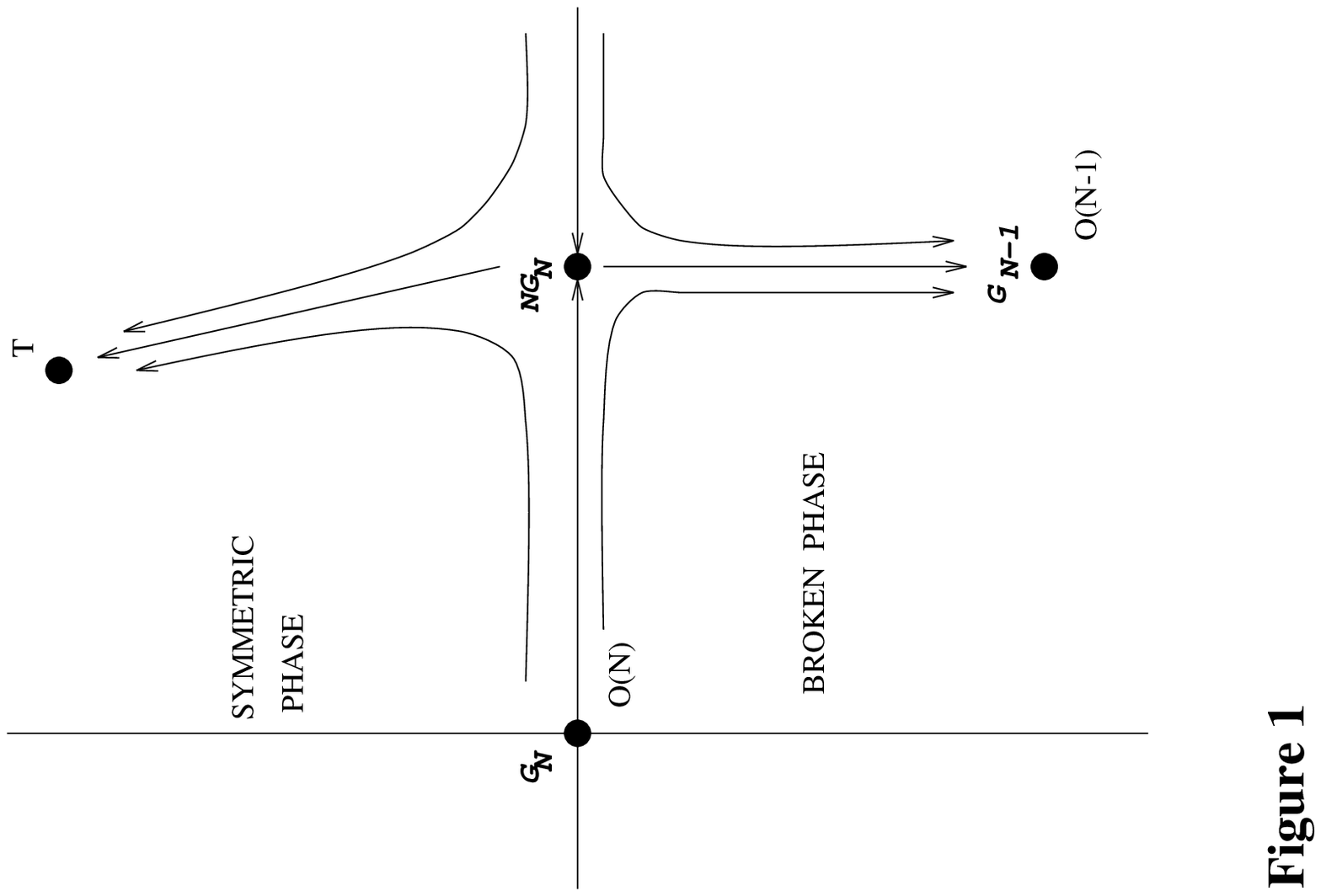,width=5truecm,height=6truecm,angle=-90}}
\caption{The fixed point structure of the O(N) sigma model; the arrows
indicate infrared flows }
\label{Fig 1.}
\end{center}
\end{figure}
At the critical point, $T=T_c, H=0$, the infrared dynamics is 
governed by the O(N) non-Gaussian fixed point ${\cal NG}_N$ when
$2<d<4$.Everywhere on the critical surface the renormalization
group (RG) trajectories flow from the the O(N) Gaussian fixed point 
${\cal G}_N$
(which is infrared unstable) to the O(N) non-Gaussian fixed point
${\cal NG}_N$.  For $T>T_c$ we have the symmetric phase with 
massive (and degenerate) 
$\sigma$ and $\vec \pi$.The infrared behaviour is governed by the 
trivial fixed point $\cal T$ which
corresponds to zero correlation length.$T<T_c$ presents a new
feature compared to N=1 case (i.e.the Ising model). The theory has
long range correlations for all $T<T_c ,H=0$ as a consequence
of the Goldstone phenomenon.Hence the infrared behaviour must be 
determined by a fixed point of
infinite correlation length.What is this fixed point? 

$T-T_c$, the deviation from the critical temperature in statistical
mechanics, is to be identified with  the square of the renormalized mass 
\cite{z} in the linear $\sigma$-model.
In particular they are parameters with positive mass dimensionality. Under 
RG flow they necessarily scale.Thus, once $T-T_c < 0$, the flow
is towards $T=0$.This means that we are interested in the zero 
temperature fixed point.  An indication as to the nature of this 
fixed point came from a 
study of $2+\epsilon$--expansion of O(N) non--linear $\sigma$--model 
by Brezin and Zinn-Justin \cite{bz}.

The O(N) non-linear $\sigma$- model in d=2 is renormalizable. 
Even though the renormalized perturbation is in terms of the
spin wave modes, Mermin-Wagner theorem \cite{z} requires that the system
is never in a phase with spontaneously broken O(N) symmetry.
A hint of this is contained in the renormalized perturbation
theory: the theory is asymptotically free.The coupling constant
grows with the distance so that the long distance structure of
the theory could be drastically different from the perturbation
theory spectrum.Indeed the spectrum is believed be that of massive 
and degenerate $\sigma$ and $\pi$'s in the
fundamental representation of O(N).

A consequence of the asymptotic freedom in $d=2$ is that in 
$d=2+\epsilon$, the $\beta$-function $\beta(t)$ has a non-trivial 
zero (say at $t=t_c$) in addition to the trivial zero, $t=0$,  corresponding 
to the theory of (N-1) free scalars.At $t=t_c$ the theory is 
scale invariant and corresponds to the critical theory 
with N massless scalars and with anomalous 
dimensions ( denoted by ${\cal NG}^{*}_N$ ).It should be noted in Fig. 1
that the fixed point ${\cal NG}_N$ of the linear $O(N)$-model is identified
with the fixed point ${\cal NG}^{*}_N$ of the non-linear model. 
This follows essentially from a remark made in \cite{z} that at ${\cal NG}_N$
the correlation functions of the linear model are identical to those of the
non-linear model where in the latter one includes correlation functions of
composite operators also. The coincidence of these two fixed points can
also be established in the $1/N$-expansion scheme \cite {zpr}.
For $t>t_c$, the coupling constant grows with the
distance, and as in $d=2$, the theory is in the unbroken phase.
$t<t_c$ is of interest to us.In this case the infrared flow
is towards $t=0$. The infrared behaviour is of O(N-1) Gaussian
theory (This fixed point is denoted
by ${\cal G}_{N-1}$ in Fig.1.)
and the ultraviolet behaviour is of the critical
theory with anomalous dimensions.

The fact that the infrared behaviour for $t<t_c$ is 
governed by the O(N-1) Gaussian fixed point gives exhaustive
information on the effects of quantum fluctuations of the 
Goldstone bosons.It means that the pions decouple from
each other at low momenta, and their propagators behave
as the free propagators for small momenta.It also allows us
to obtain the non-leading IR-behaviour 
as well as  the effects of an external magnetic field.In 
particular we now understand why various heuristic considerations
all gave the correct answer: it was the fortuitious 
behaviour of a free massless theory.Anything more 
complicated would have been much more difficult to handle.

Upto what dimension d is the analysis of $d=2+\epsilon$ calculation
correct?In particular is it applicable to d=4?The case d=4 is
somewhat different because of the absence of the non-Gaussian
fixed point ${\cal NG}_N$.At the critical temperature the infrared 
behaviour is believed to be given by free massless theory of N scalars
as the beta function is positive for small couplings; there is 
no evidence for a non-Gaussian fixed point.This appears to exclude 
the validity of the $2+\epsilon$ analysis to this case.We need 
an alternate technique to handle this situation.This is provided
,for example,by the $1/N$ expansion.

The $1/N$ expansion provides results which agree with the above analysis
for all d.(The Berlin-Kac model \cite{bk} is a variant of this technique.)
It yields the non-Gaussian fixed point for $2<d<4$ and
shows its absence for $d=4$.In either case it shows that the
leading infrared behaviour anywhere in the broken phase is given by 
N-1 free pions.The only handicap of this technique is that it 
does not provide non-anomalous dimension for the non-Gaussian 
fixed point ${\cal NG}_N$ in the leading N approximation.However even this 
is overcome once one takes higher order (in $1/N$) corrections.
There are also indications that the technique gives qualitatively
correct and quantitatively reasonable results for all $N>2$.

The fact that even in d=4 the pion dynamics is governed by the O(N-1) 
Gaussian fixed point ${\cal G}_{N-1}$ in the infrared has important 
implications.In particular the propagator of $\sigma$ 
(to be interpreted as the O(N) partner of the pions)
has an universal logarithmic singularity in the infrared region,

\begin{equation}
\Delta_{\sigma} (k) \sim ln(\frac{\mu}{k^2})
\end{equation}

\noindent where $\mu$ is the renormalisation scale.
This follows from the same arguments as for $d<4$, in particular 
because $\sigma \sim :\bf \pi^2:$ in the infrared.This feature
is present whether $\sigma$-propagator  has a pole at 
some $k^2 \neq 0$ or not. The precise location of the pole is 
a non-universal feature depending on the microscopic 
dynamics.  It could have interesting implications for
nuclear forces at low energies in the 'sigma'-channel (i.e., 
spin J=0,isospin I=0) as will be discussed elsewhere.

In Sec. 3 and 4 we develop a new formalism for $N 
\rightarrow \infty$ which will allow for explicit 
calculation of all correlation functions even away
from the infrared regime.We will
explicitly demonstrate all the properties mentioned above.Our 
formalism gives the same results in the leading infrared limit whether
one starts with the linear or the non-linear $\sigma$- model.

That the pions behave like free massless bosons in the 
infrared limit, in any $d>2$, is such a general phenomenon that
there should be an easier way of understanding it.
A careful analysis of the proof of the Goldstone theorem
along with current algebra and PCAC provide one such  
non-perturbative argument.Another is the treatment based on Ward
identities given in sec 5.Low energy theorems based on PCAC and
current algebra do imply that 
pions  decouple from each other and with other matter at low
momenta, but they  do not directly imply that the pion has a
canonical dimension in the infrared.Current algebra requires that the
current coupling to the pion has canonical dimension $d-1$. 
However, the canonical dimension (for the $k \rightarrow 0$
behaviour of the pion) is $(d-2)/2$.
This only means that the PCAC relating the two needs a
dimensionful parameter, but does not directly relate the
two dimensions.

Consider the time ordered correlation function of a
current in a broken direction and the corresponding $\pi$ field:

\begin{equation}
\Delta^{ii}_{\mu}(x) = 
<0 \mid T(j_{\mu}^i(x), \pi^i (\bf 0)) \mid 0>
\end{equation}
We get,

\begin{equation}
\int d^dx \partial^{\mu} \Delta^{ii}_{\mu}(x) = 
<0 \mid  [Q^i, \pi^i (0)] \mid 0> \not = 0
\end{equation}
since $ [Q^i, \pi^i (0)] = \sigma(0) $ has a non-zero expectation value.
This means for the Fourier transform of this correlation
function,

\begin{equation}
lim_{q \rightarrow 0} q^{\mu} \Delta^{ii}_{\mu}(q)  \neq  0
\end{equation}
This requires that in the spectral representation of the
correlation function there is necessarily a 
$\delta(m^2)$ contribution implying a {\em simple pole} 
at zero mass.Any other singularity, such as a branch point
will not satisfy the above equation.

The fact that the $\sigma$ channel also has an
infrared  singularity is equally general and should be
understandable in a general way.
In Sec. 5 and 6 we provide two proofs which rely  essentially
on symmetry arguments. The first uses Ward identities, 
and the second, the Schwinger- Dyson equations. 

\section{$\Gamma$ in $N \rightarrow \infty$ limit}

It was argued in Sec. 2 that the leading term in $1/N$ expansion
retains all important effects of the Goldstone boson fluctuations.
Analytic calculations are possible in this limit and therefore 
we can analyse such effects in detail.There is an extensive
literature on this technique \cite{z}.In this section we 
provide a new arsenal to this technique.We define a generalized
generating functional of one-particle irreducible (1PI) vertices 
and obtain an  explicit expression for it in the large N limit.
This is made possible by involving an  auxiliary field.
Our method provides 
the easiest way of extracting all correlation functions
without doing a separate calculation for each \cite{amit}. 
We present results
for both linear and non-linear $\sigma$-models.

$W[\vec H]$, the generating functional of connected Green's
functions, is given by,

\begin{equation}
e^{i W[\vec H]} = \int {\cal D} {\vec \Phi} 
e^ {i(S[{\vec \Phi}] + {\vec H}.{\vec \Phi})}
\end{equation}
where,
\begin{equation}
\vec H.\vec \Phi= \int d^dx  \vec H(x).\vec\Phi(x)
\end{equation}
Connected Green's functions of $\Phi_i$ are obtained from 
various functional derivatives of $W[\vec H]$.
$\Gamma[\vec \Phi]$, the generating functional of 1PI vertices, 
is obtained from $W[\vec H ]$ by a Legendre transformation.

\begin{equation}
\Gamma[\vec \Phi] =   W[\vec H] - \vec H.\vec \Phi
\end{equation}

where, 

\begin{equation}
\Phi_i(x)= \frac {\delta W[\vec  H]}{\delta H_i(x)}
\end{equation}

The vacuum expectation value (VEV) of $\Phi_i$
is obtained as the value at which $\Gamma$  
is minimized.The second functional derivative about this minimum
gives the inverse propagator. Further, various functional 
derivatives of $\Gamma$ about this minimum give 
the 1PI vertices. The connected Green's functions
can be extracted by building {\it only} tree diagrams using these
propagators and vertices.

In the linear $\sigma$-model, introduce an 
auxiliary field $\lambda$ to make the exponent bilinear in $\vec \Phi$:

\begin{equation}
e^{i W[\vec H]} = \int {\cal D} \lambda {\cal D} {\vec \Phi}
e^{i \int d^dx (\frac{1}{2} (\partial_{\mu} {\vec \Phi}(x))^{2}
-\frac{1}{2} \lambda(x)({\vec \Phi}^2(x)-C^2)
+\frac{1}{4U} \lambda^{2}(x)+{\vec  H}(x) \cdot {\vec \Phi}(x)}
\end{equation}
Now an integration over each
$\Phi_i, i=1,2....,N$, gives a $(Det)^{-1/2}$ factor.
Therefore,

\begin{equation}
e^{i W[\vec H]} = \int {\cal D}  \lambda 
e^{i \int d^dx (\frac{1}{4U} \lambda^{2}(x)+
\frac{C^2}{2} \lambda (x))
-\frac{N}{2} Tr ln \nabla -i \frac{1}{2} H.\nabla^{-1}.H}
\end{equation}

where,
\begin{equation}
\nabla (x,y) = ( -\partial^2 - \lambda(x)) \delta
(x-y)
\end{equation}

On defining,

\begin{equation}
NU = u, C^2 = N c^2,
H_i = \sqrt N h_i,
W[{\vec H}] = N w[{\vec h}] 
\label{redefs}
\end{equation}

We get,

\begin{equation}
e^{i Nw[{\vec h}]} = \int {\cal D}  \lambda 
e^{i N [(\int d^dx(\frac{1}{4u} \lambda^{2}
+\frac{1}{2}c^2 \lambda))
+ \frac{i}{2}Tr ln \nabla - \frac{1}{2} h. \nabla^{-1}.h]}
\end{equation}

We are interested in the limit $N \rightarrow \infty $,
$U \rightarrow 0$ holding  $NU = u$ fixed. With the variables 
redefined as in Eqn. \ref{redefs}, it is clear that this limit is dominated 
by the saddle point. The fluctuations about the saddle point
give contributions which are non-leading in N.
Thus the leading contribution is,

\begin{equation}
w_{\infty}[{\vec h}] = 
\int d^dx(\frac{1}{4u} \lambda^{2}(x)+\frac{1}{2}c^2 \lambda(x)) \\
+\frac{i}{2} Tr ln \nabla - \frac{1}{2} h. \nabla^{-1}.h
\label{ggf2}
\end{equation}
where $\lambda(x)$, the saddle point, is a functional of
$\vec h(x)$ as given by,

\begin{equation}
\frac{1}{u} \lambda(x)+c^2 
-i  \nabla^{-1}(x,x) 
- (h. \nabla^{-1})(x).(\nabla^{-1}.h)(x) = 0
\label{lam}
\end{equation}

We may obtain $\Gamma[\vec \Phi]$ from $w_{\infty}[{\vec h}]$ by
a Legendre transform as described earlier. But this form is
not easily amenable to further analysis. Instead we note 
that the saddle point equation  may be rewritten as  

\begin{equation}
\frac{1}{u} \lambda(x)+c^2 
-i \nabla^{-1}(x,x) 
=\vec \phi^2(x) 
\label{saddle}
\end{equation}
where,
\begin{equation}
\nabla \phi_i(x)= -h_i(x) 
\end{equation}
Notice that these equations can be obtained from the following 
functional of $\vec \phi$ and $\lambda$ :

\begin{eqnarray}
\Gamma_{\infty}[\vec \phi,\lambda] & = & 
\int d^dx(\frac{1}{4u} \lambda^{2}(x)
+\frac{1}{2}c^2 \lambda(x)
+ \frac{1}{2}(\partial_{\mu} \vec \phi)^2 (x) \nonumber \\  
&&- \frac{1}{2} \lambda(x) \vec \phi^2(x))
+ \frac{i}{2}Tr ln \nabla  
\label{gammainf}
\end{eqnarray}
by taking functional derivatives:

\begin{equation}
-h_i(x)= \frac {\delta \Gamma_{\infty} [\vec \phi,\lambda]}
{\delta \phi_i(x)}, \quad
- \kappa(x)=\frac{\delta \Gamma_{\infty}[\vec \phi,\lambda]}  
{\delta \lambda(x)}    \label{ggf1}
\end{equation}
Here we have temporarily added a 'source' $\kappa$ for $\lambda$.When
restricted to correlation functions with only $\phi_i$ on external legs,
$\kappa$ can be set to zero, yielding eqn 30.
These equations imply that 
$w_{\infty}[\vec h,\kappa]$ and 
$\Gamma_{\infty}[\vec \phi,\lambda]$  
are related by a functional Legendre transformation in both
arguments.
This means that the connected Green's functions
of $\vec \phi$ and in fact also of  $\vec \phi^2$ (i.e. functional
derivatives of $w_{\infty}[\vec h,\kappa]$  wrt both $\vec h $ and $\kappa$)
can be obtained from tree diagrams built from the 'action'
$\Gamma_{\infty}[\vec  \phi,\lambda]$.This connection  is simply
an algebraic consequence of the Legendre transformation.  
We have thus the option of constructing Green's functions involving
the composite operator $:\phi^2:$ also by having $\lambda$  as
external legs.

By introducing an auxiliary field $\lambda$ we have obtained
an explicit form of the generating functional of 1PI vertices
which is far more amenable to analysis than if we had obtained a 
$\Gamma(\vec\phi)$ by substituting the saddle point solution for 
$\lambda$ .Moreover we find that
the properties of the phase with spontaneously broken symmetry
can be understood in a simple way in terms of this auxiliary field.
Ofcourse,
now $\lambda$ also appears in propagators and vertices.
The usual generating function $\Gamma[\phi]$
can however be recovered by using  $\kappa =0$ from Eq.\ref{ggf1} 
to replace $\lambda$ as a functional of
$\phi$ in $\Gamma_{\infty}[\vec \phi, \lambda]$.

There is hardly any change in our calculations (and results) when
passing over to the 
non-linear $\sigma$-model.
Now we have,
\begin{equation}
e^{i W_{NL}[\vec  H]} = \int {\cal D} \vec \Phi \Pi_x
\delta^d(\vec \Phi^2(x)-C^2)
e^{i \int d^dx (\frac{1}{2} (\partial_{\mu} 
\vec \Phi(x)^{2} +\vec H(x) \cdot \vec \Phi (x)})
\end{equation}
Using the Fourier representation for the functional 
$\delta$ function,
\begin{equation}
e^{i W_{NL}[{\vec H}]} = \int {\cal D} \lambda {\cal D} \vec \Phi
e^{ i\int d^dx (\frac{1}{2} 
(\partial_{\mu} \vec \Phi(x)^{2})
-\frac{1}{2} \lambda (\vec \Phi^2 (x)-C^2)
+\vec  H(x) \cdot \vec \Phi(x))}
\end{equation}

In this form the only difference with the linear $\sigma$-model
is that a quadratic term in $\lambda$
is missing in the exponent.Equivalently, the 
non-linear $\sigma$-model is simply the special case
$U \rightarrow \infty$ of the linear $\sigma$-model.
Thus all results for the non-linear $\sigma$-model
can be extracted by taking this limit in our formulae.We find
that it is straightforward to take this limit and the results 
are not very different.

We have been cavalier in our manipulations with the functional
integrals.For instance, the $Tr ln$ term has to be carefully
defined.We may choose a Pauli-Villars or a lattice cut-off
for defining this term.If we consider a functional Taylor
expansion of the Tr ln term about a constant value of 
$\lambda$, for $2<d<4$,  only the term linear in $\lambda$
is sensitive to the cut-off.We may therefore obtain a renormalized 
$\Gamma_{\infty}[\bf \phi,\lambda]$  
by making a specific cut-off independent choice for this term.
In $d=4$, in addition,  the term quadratic in $\lambda$ is also
divergent and requires a renormalization.For the sake of uniformity,
we shall introduce local counterterms $a \lambda +b \lambda^2$ for
both $d=3$ and $d=4$. In the $d=3$ case, this amounts to a finite
renormalisation of u. Because of the presence of massless modes,
renormalisation should be so performed that it does not introduce
spurious infra-red divergences. Hence we choose to define renormalised
quantities at $\lambda=\mu$(say). Then the renormalised effective action
is
\begin{eqnarray}
\Gamma_{\infty}[\vec \phi,\lambda] & = & 
\int d^dx(\frac{1}{4u}( \lambda(x)-\mu)^{2}
+\frac{1}{2}c^2 (\lambda(x)-\mu)
+ \frac{1}{2}(\partial_{\mu} \vec \phi)^2 (x) \nonumber \\  
&&- \frac{1}{2} \lambda(x) \vec \phi^2(x))
+ \frac{i}{2}Tr^\prime_{\mu} ln \nabla  
\label{gammaren}
\end{eqnarray}
$Tr^\prime_{\mu} $ means that the {\bf local} part of the first  two 
terms in the functional expansion around $\lambda=\mu$
are to be ignored.Now u and c are renormalised quantities.
Thus the model is renormalizable in the usual sense.It is
also seen that the wave function renormalization  for $\lambda$,
equivalently the composite operator $\vec \phi^2$,
is finite in the
$N \rightarrow \infty $ limit.(For a discussion of renormalizability
of the $1/N$ series see Ref. \cite{russ}).

The discussion of renormalizability is equally valid for the
the non-linear $\sigma$-model.  The only difference is that 
the bare $ \lambda^{2}(x)$ term is
not present in the action.The quantum effects introduce such
a term which is finite in $2<d<4$ and diverges logarithmically
with the cut-off in $d=4$.Thus the non-linear $\sigma$-model
is not renormalizable in this sense in $d=4$ even in the $N
\rightarrow \infty$ limit.

\section{Correlation functions in $N \rightarrow \infty$ limit}

In this section, we extract the infrared behaviour of the 
correlation functions, in particular, in the phase with SBS.

First we need to obtain the VEV's $v_i$ and $\lambda_0$
of the fields $\phi_i$ and $\lambda$.For uniform (in particular
zero) fields $\vec  h$ (and $\kappa$), the VEV's are space-
time independent.Therefore we get

\begin{equation}
\lambda_0 \vec  v =\vec  h
\end{equation}
\begin{equation}
\frac{1}{u} (\lambda_0-\mu) = (v^2 - c^2)+
i \int \frac{d^d k}{(2\pi)^d}
[\frac{1}{k^2- \lambda_0}-\frac{1}{k^2-\mu}-
\frac{\lambda_0-\mu}{(k^2-\mu)^2}]   \label{mini}
\end{equation}

The analysis below closely follows Ref. \cite{z}.
We first consider the situation in the absence of the external magnetic
field, $\vec h = 0$.Now $\lambda_0 v = 0$ and so there are three cases:
\begin{enumerate}
\item[i)]
$\lambda_0=0$ and $v=0$ corresponds to the critical point.This is 
reached for the critical value $c=c_{cr}$ where,

\begin{equation}
c_{cr}^2 = \frac {\mu}{u}+i\mu^2 \int \frac{d^d k}{(2\pi)^d} 
\frac{1}{k^2(k^2-\mu)^2}
\label{casei}
\end{equation}
\item[ii)]
$ \lambda_0 \neq 0$ and $ v=0$ is the symmetric phase.
$\lambda_0$ is given by,

\begin{equation}
\lambda_0 (\frac{1}{u}
-i  \int \frac{d^d k}{(2 \pi)^d} \frac{\mu^2+\lambda_0 k^2
-2\mu k^2}{k^2 (k^2- \lambda_0)(k^2-\mu)^2})
=(c_{cr}^2 - c^2)
\label{caseii}
\end{equation}
where we have used Eqn.\ref{casei} 
\item[iii)]
$ \lambda_0=0$, $ v \neq 0$ is the phase with SBS.The VEV
$v$ is given by,

\begin{equation}
v^2=(c^2 - c_{cr}^2)
\end{equation}
as seen by using Eqn.\ref{casei}
\end{enumerate}
For any $c^2 > c_{cr}^2$ we have a solution for a real $v$
and this corresponds to a phase with SBS.
On the other hand for any $c^2 \leq c_{cr}^2$
we have a solution for $\lambda_0$ in Eqn.\ref{caseii}, corresponding to
the symmetric phase.

In terms of the fluctuations
\begin{equation}
\lambda'(x) = \lambda(x)- \lambda_0
\end{equation}
we have for the symmetric phase,
\begin{eqnarray}
\Gamma_{\infty}[\vec \phi,\lambda]
&=& \int d^dx (\frac{1}{2} (\partial_{\mu} \vec \phi(x)^{2})
-\frac{1}{2}\lambda_0 \vec \phi^2 (x)
-\frac{1}{2}\lambda'(x) \vec \phi^2 (x)) \nonumber \\
& & +\frac{1}{4} \int \int d^dx d^dy \lambda'(x)I(x-y,\lambda_0) \lambda'(y) 
\nonumber \\
& & + \frac{i}{2}Tr'_\mu ln (  -\partial^2- \lambda_0-\lambda')
\end{eqnarray}
Where $I(x-y,\lambda_0)$ is the fourier transform of
\begin{equation}
I(p^2,\lambda_0)=i\int d^dk [\frac{1}{(k^2-\mu)^2}-\frac{1}{(k^2-\lambda_0)
((k+p)^2-\lambda_0)}]+\frac{1}{u}
\end{equation}

All terms linear in the fluctuations drop out because
we are expanding about the extremum value.$\lambda_0$ is seen to
be the $(mass)^2$ of the $\phi$ field.

For the phase with SBS we choose 
$v_N =v \neq 0,v_i = 0, i=1,...(N-1)$. 
In terms of the fluctuations,

\begin{equation}
\sigma(x) = \phi_N(x)-v
\end{equation}
we get,
\begin{eqnarray}
\Gamma_{\infty}[\vec\Phi,\sigma,\lambda]
&=& \int d^dx (\frac{1}{2} (\partial_{\mu} \vec \phi(x))^{2}
-v \lambda(x)\sigma(x) -\frac{1}{2} \lambda(x) (\sigma^{2}(x)+\vec \pi^2(x))
\nonumber \\
& & +\frac{1}{4u} \int \int d^dx d^dy \lambda (x)I(x-y,0) \lambda (y) 
+i \frac{1}{2}Tr'_\mu ln (  -\partial^2-\lambda)\nonumber \\
\end{eqnarray}
$\vec  \pi$ is massless as is to be expected. The mixing of
$\sigma$ with the auxilary field $\lambda$ is a crucial
feature of this phase. This mixing governs the infrared
behaviour of the correlation functions as seen below.

The terms quadratic in $\sigma$ and $\lambda$ may be
written in momentum space as,

\begin{equation}
i( \frac{1}{2} \lambda(- k) I(k^2,0) \lambda(k)
-v \lambda (- k) \sigma(k)
+ \frac{1}{2} \sigma(- k) k^2 \sigma (k)) \label{quad}
\end{equation}
Now,
\begin{eqnarray}
I(k^2) \equiv I(k^2,0)&=& \frac{1}{2u}- (k^2)^{d/2-2},~~ 2<d<4,   \nonumber \\
&  &\sim  \frac{1}{2u}+ln(\mu/ k^2), ~~ d=4
\end{eqnarray}
We get the propagators by inverting the matrix of the quadratic 
form in Eqn.\ref{quad}.
 
\begin{eqnarray}
<\sigma(- k) \sigma( k)>
&=& \frac{i}{k^2-v^2/{I(k^2)}}  \nonumber \\
<\sigma(- k) \lambda( k)>
&=& \frac{iv}{k^2 I(k^2)-v^2}  \nonumber \\
<\lambda(- k) \lambda(k)>
&=& \frac{ik^2}{k^2 I(k^2)-v^2}
\end{eqnarray}

We are now in a position to extract the infrared behavior 
of various correlation functions.
The propagators have the following infrared behaviour:
\begin{eqnarray}
<\pi_i(-k) \pi_j(k)> & \sim &
\frac{i\delta_{ij}}{k^{2}},   \nonumber \\
<\sigma(-k) \sigma(k)> & \sim & 
\frac{i}{\mid {k}\mid^{4-d}},   \nonumber \\
<\sigma(-k) \lambda(k)> & \sim &  \frac{i}{v},   \nonumber
\\
<\lambda(-k) \lambda(k)> & \sim &  ik^2
\end{eqnarray}

Thus the leading N approximation  reproduces the correct 
infrared behaviour for the $\pi$ and the $\sigma$ propagators,
viz., $\pi$ has canonical $k^{-2}$ behaviour, whereas $\sigma$
has an anomalous dimension $(d-2)$, precisely as if $\sigma
\sim :\bf \pi ^2:$.We now demonstrate explicitly that the
other correlation functions also have the infrared
behaviour required by the soft pion theorems.

The interactions of $\pi$ and $\sigma$ are only
mediated through the auxiliary field $\lambda$ in this
formalism, and the  $\lambda$ propagator has an unusual $k^2$
infrared behaviour.This is the essential reason why $\pi$ 's and $\sigma$ 's
decouple from each other in the infrared, and the results 
of soft pion theorems are valid.

For the leading IR-behaviour, the multi-$\lambda$ vertices are 
irrelevant. The effective action is then  quadratic
in $\lambda$ and consequently, the $\lambda$-field can be exactly integrated
out.This leads to the following effective action to describe the leading IR
behaviour:
\begin{equation}
\Gamma^{IR}=\int d^dk[\frac{k^2}{2}|\vec\phi(k)|^2+\frac {|(\vec \phi^2
+2v \sigma)(k)|^2}{4I(k^2)}]
\end{equation}
This summarises all the essential results of this paper. 

Thus the large N approximation provides a valuable non--
perturbative tool for obtaining effects due to quantum
fluctuations of Goldstone bosons.It incorporates the
infrared effects correctly.It also incorporates the 
ultraviolet behavior correctly: for $2<d<4$, the 
ultraviolet behaviour is of the non-Gaussian fixed point,
and for d=4, it is of the O(N) Gaussian fixed point. 
It has an explicit $\sigma$ field, so that effects due to its infrared 
behaviour on other matter can be investigated. 
Also the technique provides a way of going beyond the leading infrared
region.It provides a detailed structure for the propagators of
both $\pi$ and $\sigma$ with correct asymptotic behaviours.

We now consider the situation with a uniform external magnetic field
$\vec h=(0,0,......,h)$.
Now, $\lambda_0 v = h$ implies that neither $\lambda_0$ nor $v $ 
is zero.This is the choice that removes terms linear in the shifted
field $\sigma$  one gets essentially the same $\Gamma_{\infty}$ 
as the one we got for the symmetric phase but with a crucial additional
term $-v\lambda'(x) \sigma(x)$ and with the proviso that elsewhere
$\vec \phi^2$ now stands for $\sigma^2+\vec\pi^2$.

Thus $\pi$ has a mass $\sqrt \lambda_0$. $\sigma$ has a mixing with
$\lambda$ as in the SBS case.The $\sigma \lambda$ propagators may
be computed as before.

The equation governing $\lambda_0$ is now,

\begin{equation}
\frac{1}{u} \lambda_0 = (v^2 - \delta c^2)+
i\lambda_0\int \frac{d^d k}{(2\pi)^d}
(\frac{1}{k^2(k^2- \lambda_0)} - \frac{1}{(k^2-\mu)^2})
\label{state}
\end{equation}
with $\delta c^2 = c^2-c_{cr}^2$.
By an analysis of these equations one obtains a mass
that scales as $h^{1/2}$ for the pions and as $h^{1-d/4}$ for the
sigma.

For applications to statistical mechanics, one should use the expressions
derived here continued to the relevant number of euclidean dimensions.

\section{Use of Ward identities}
In this section we demonstrate that the $\chi_L$ 
diverges for $2<d \leq 4$  ,that $\sigma \sim \vec \pi^2$
in the infrared regime  and the softness of multipion scattering
amplitudes using Ward identities. 

Chiral Ward identities merely reflect the
underlying symmetry structure of the theory and are useful both 
in the symmetric phase where they relate correlation functions
of the same type ( n-point functions with fixed n) as well as in the
ordered phase where they relate different n-point functions in a
non-trivial way. 
Note that in the generating function 
both the measure ${\cal D} \vec \phi$, and the action $S(\vec \phi)$, are
invariant under $O(N)$ rotations generated by $T_a$;the 
source term is also  invariant under a simultaneous rotation of both the
source $\vec H$ and the $\vec \phi$ field.Considering an 
infinitesimal transformation, this implies for $\Gamma$, 

\begin{equation}
\int d^dx [ \phi_i(x) ({\delta \Gamma \over \delta \phi_j
(x)}-H_j)-\phi_j(x) ({\delta \Gamma \over 
\delta \phi_i(x)}-H_i)] =0  \label{B}
\end{equation}
when an external field $\vec H$ is present.This means that $\Gamma$
has only O(N) invariant combinations such as 
$\vec \phi^2,( \partial_{\mu} \vec \phi)^2$ etc., except for one additional
term $\int \vec H . \vec \phi$, depending on the external field.

By successive differentiations of this identity w.r.t. the fields
$\sigma,\vec\pi$, one can generate various forms of Ward Identities.In
particular, an identity that will prove most useful later on is
\begin{equation}
\Gamma^{(2)}_{\sigma \sigma} (p) - \Gamma^{(2)}_{\pi\pi}(p) =
<\sigma> \Gamma_{\pi\sigma\pi} (p,0,-p) \label{A}
\end{equation}
where $<\sigma>$ is the expectation value of the order parameter.\\ 

To explore the infrared properties,  we consider the thermodynamic 
potential, which is called the the exact effective potential in quantum field 
theory.  We include an external magnetic field to serve as an 
infrared regulator and consider the singular behaviour as this field
goes to zero.  The various zero momentum n-point functions
are obtained by functionally differentiating this potential w.r.t.
constant fields and evaluating the derivatives at the minima of the
potential. In the ordered phase this minima occurs at a non-vanishing
value $<\sigma>=v$ as the magnetic field is reduced to zero.\\

As a consequence of the Ward identity, Eqn.\ref{B},
the effective potential has the form,

\begin{equation}
U(\vec\phi ) = f(\vec\phi \cdot \vec\phi) + \vec H \cdot \vec \phi
\end{equation}
Since
\begin{equation}
U_i \equiv {\delta U\over{\delta\phi_i}} = 2\phi_i f^{\prime}-H_i
\end{equation}
the VEV $v$ is given by,
\begin{equation}
f^{\prime}(v^2) = {H\over {2v} }
\end{equation}
where H is the magnetic field along the $\sigma$-direction. 
We also have
\begin{equation}
U_{ij} \equiv {\delta^2 U\over{\delta\phi_i
\delta\phi_j}}=2\delta_{ij}f^{\prime} + 4\phi_i \phi_j f^{\prime\prime}
\end{equation}
\begin{equation}
U_{ijk} \equiv 4(\delta_{ij}\phi_k + (cyclic))f^{\prime\prime} +
8\phi_i\phi_j\phi_k f^{\prime\prime\prime}  \label{E}
\end{equation}
\begin{equation}
U_{ijkl} = 4(\delta_{ij}\delta_{kl} +(cyclic))f^{\prime\prime} +
8(\delta_{ij}\phi_k\phi_l +(cyclic))f^{\prime\prime\prime} +
16\phi_i\phi_j\phi_k\phi_l f^{iv} \label{F}
\end{equation}
etc. Here $f', f'' $ etc. are the derivatives of $f$ w.r.t. its
argument.  Since
\begin{equation}
U_{\pi_i \pi_j}\vert_{min}=m_{\pi}^2 \delta_{ij}
\end{equation}
we get,
\begin{equation}
m_{\pi}^2=2f'(v^2)={H\over v} \label{D}
\end{equation}
Also,
\begin{equation}
m_{\sigma}^2= 2 f^{\prime}(v^2) + 4 v^2 f^{\prime\prime}(v^2) \label{C}
\end{equation}
It can be seen that Eqns. \ref{D}, \ref{C} and \ref{E} together are equivalent 
to the $k \rightarrow 0$ limit of the  identity in Eq.\ref{A}. 
>From Eq.\ref{F},the 
1PI pion four - point vertex at zero momentum is given by
\begin{equation}
U_{4\pi} = 4(\delta_{ij} \delta_{kl}+ cyclic) f^{\prime\prime}(v^2)
\end{equation}
and from Eq.\ref{E}, the $\sigma\pi\pi$ 1PI vertex is given by
\begin{equation}
U_{\sigma\pi_i\pi_j}= 4v \delta_{ij} f^{\prime\prime}(v^2)
\end{equation}
Hence the $\pi-\pi$-scattering amplitude at zero momentum  \cite{wein} is,
\begin{equation}
A_{4\pi}=i(\delta_{ij}\delta_{kl}+cyclic..)[4 f^{\prime\prime}
-{16 v^2 (f^{\prime\prime}(v^2))^2\over {2 f^{\prime}(v^2) + 4
v^2 f^{\prime\prime}(v^2)}}]
\end{equation}
In other words
\begin{equation}
A_{4\pi}={4i H f^{\prime\prime}(v^2)\over{H + 4v^3
f^{\prime\prime}(v^2)}} (\delta_{ij}\delta_{kl}+ cyclic)
\end{equation}
Irrespective of $f^{\prime\prime}$, $A_{4\pi}$ is at least as soft as H.
This is equivalent to the soft pion theorems of current algebra \cite{spt}.
Note that this result has been proven here without the detailed
assumptions of current algebra and PCAC.\\

On introducing the shorthand notation
\begin{equation}
I_{i_1i_2...i_n} =
2^{n/2}[\delta_{i_1i_2}\delta_{i_3i_4}....\delta_{i_{n-1}i_n}+ cyclic]
\end{equation}
and
\begin{equation}
\phi_{i_1i_2...i_n} = 2^n\phi_{i_1}\phi_{i_2}....\phi_{i_n}
\end{equation}
one can write the general result
\begin{equation}
U_{i_1i_2..i_N} = \sum_{\stackrel{k,l}{2k+l=N}}
	I_{i_{l+1}i_{l+2}....i_{l+2k}} \phi_{i_1...i_l}f^{(k+l)}+cyclic
\end{equation}

Now let us consider the zero momentum 1PI vertex function
$\Gamma_{n\sigma m\pi}$ (with $N=n+m$) corresponding to n-longitudinal 
and m-transverse
external legs.Then all tensors $\phi_{i_1...i_l}$ will vanish when $l>n$
as then at least some of the $\phi$'s in these tensors will have to be
transverse.Furthermore the leading IR behaviour is equivalent to treating
$v$ as being very large (compared to the momenta and
equivalently to h ). It is clear that due to the exchange of the massless
goldstone particles in the intermediate state, the n-point functions should
have infrared singularities for sufficiently large $n$. From very general
considerations, it follows that the nature of these singularities becomes
more severe with increasing $n$. Thus, for sufficiently large $n_0$, the
derivative $f^{n_0}(\sigma^2)$ has the structure 
\begin{equation} 
f^{(n_0)}(\sigma^2)=\sum_{k>0}C_k(f^\prime(\sigma^2))^{-k}+C_\alpha
+\sum_{l>0}C_l(f^\prime(\sigma^2))^l
\end{equation}
where
\begin{equation}
f^\prime(\sigma^2)\vert_{\sigma^2=v^2}=\frac{H}{v}
\end{equation}
Therefore
\begin{equation}
f^{(n_0+1)}(\sigma^2)=-\sum_{k>0}kC_k(f^\prime(\sigma^2))^{-k}
\frac{f^{\prime\prime}(\sigma^2)}{f^\prime(\sigma^2)} +\sum_{l>0}C_l l
(f^\prime(\sigma^2))^l\frac{f^{\prime\prime}(\sigma^2)}{f^\prime(\sigma^2)}
\end{equation}

Now three situations can arise: (a) $f^{\prime\prime}(\sigma^2)$ is finite 
or goes to zero slower than $f^\prime(\sigma^2)$ as $\sigma\rightarrow v$.
Then $f^{(n_0+1)}$
is more singular than $f^{(n_0)}$. Also $A_{4\pi}\rightarrow H$ from (67):
(b)$f^{\prime\prime}(\sigma^2)$ and $f^\prime(\sigma^2)$ have the same 
leading infrared
behaviour. Further, (63) already implies that $m_\sigma^2$ and $m_\pi^2
\sim \frac{H}{v}$ which is a rather strong result.
Then all $f^{(n)}$ for $n>n_0$ have the same infrared behaviour. We
shall show shortly that this is inconsistent. 
(c) $f^"(\sigma^2)$ vanishes faster than $f^\prime(\sigma^2)$ as $H\rightarrow
0$. This has the consequence that $f^{(n_1)}$ has softer IR singularity than
$f^{(n_2)}$ for $n_1>n_2$. By virtue of (63) this choice also implies that 
$m_\sigma^2 = m_\pi^2 = \frac{H}{v}$. This choice also implies that  
$A_{4\pi}$ vanishes faster than $H$ from (67). As these features are
already invalidated in perturbation theory, (c) will not be considered
any more. 

Now, for both the situations a) and b) it follows that the dominant
contribution to $\Gamma_{n\sigma m\pi}$
will come from $l =n\ (k=m/2)$ in (70); this is the maximum possible
value for $l$. Hence
\begin{equation}
\Gamma_{n\sigma m\pi}~~~\sim f^{(n+m/2)}
\end{equation}
This further implies that
\begin{equation}
\Gamma_{n\sigma m\pi}~~~\sim \Gamma_{(2n+m)\pi}
\end{equation}
This means that as far as the
infrared behaviour of all the vertex functions is concerned 
\begin{equation}
\sigma~~~~\sim~~\pi^2
\end{equation}
This and the fact that pions are non-interacting in the IR 
leads to the stated result on  $\chi_L$ {\em viz.} $\chi_L \sim 
H^{\frac{d-4}{2}}$ for $2<d<4$ and $\chi_L\sim log H$ for $d=4$.

But these behaviours for $\chi_L$ are inconsistent with the situation (b)
which demand $\chi_L\sim H^{-1}$. Thus only (a) need be considered. 

In order to get an appreciation for the  nature of the effective 
potential,we analyse it in the large N limit.From the
generating functional $\Gamma[\vec\phi,\lambda]$ of Sec.3.
{}one can recover the usual generating functional
$\Gamma[\vec\phi]$ by eliminating $\lambda$ using its equation
of motion. Since we are interested in constant fields 
$\vec\phi$, it is sufficient to have the $\lambda$ field also constant.
Subtracting Eqn.~\ref{mini} from the analog of Eqn.~\ref{saddle} that results
from the renormalised $\Gamma_{\infty}$(eqn \ref{gammaren}), we get,
\begin{equation}
(\lambda -\lambda_0)[\frac{1}{u}-i \int \frac{d^d{k}}{(2\pi)^d}
(\frac{1}{(k^2-\lambda)(k^2-\lambda_0)}-\frac{1}{(k^2-\mu)^2})]
=(\vec\phi^2 - v^2)
\end{equation}
Which can be solved to yield the implicit solution
\begin{equation}
\lambda = f(\vec\phi^2 - v^2)
\label{la}
\end{equation}
for given $\lambda_0,u$.Therefore the effective potential has the form,
\begin{equation}
U_{\infty}(\phi)=
\frac{ (\lambda-\mu)^2}{4u} -\frac{ \lambda (\vec\phi^2-c^2)}{2}
+\frac{i}{2} Tr^\prime_\mu ln \nabla
+\vec h \cdot \vec\phi
\end{equation}
where $\lambda$ is to be substituted as in Eqn.~\ref{la}.
Further we know that the minimum of the effective potential is 
at, $\phi^2=v^2, \lambda=\lambda_0$ such that $\lambda_0
\rightarrow 0, v \neq 0,\  v \lambda_0 =h$, as 
$ h \rightarrow 0$.This means that $\lambda$ and also 
$U_{\infty}(\vec\phi^2)$ when expanded about the minimum
$\vec\phi^2=v^2$ has Taylor coefficients which are increasingly
singular in $\lambda_0$ and hence in $h$.

\section{Use of Schwinger - Dyson Equations}

In this section we present an argument based on 
Schwinger - Dyson equations that shows the diverging $\chi_L$
as a consequence of the
infrared divergence of the Goldstone bosons.

The heirarchy of Schwinger-Dyson equations are obtained by exploiting
the following property of functional integrals
\begin{equation}
0= \int {\cal D}\phi {\delta \over {\delta \phi_i(x)}}
e^{i(S+J \cdot \phi)} 
\end{equation}
To be specific let us consider
\begin{equation}
S(\vec \phi) = \int d^dx
[\frac{1}{2}\partial_\mu\vec\phi\cdot\partial^{\mu}\vec\phi
-\frac{1}{2}\mu^2\vec\phi\cdot\vec\phi - 
\frac{1}{24N}\lambda(\vec\phi\cdot\vec\phi)^2 ]
\end{equation}
The above identity implies the following equation for
$\Gamma$, the generating functional of 1PI diagrams:
\begin{eqnarray}
{\delta \Gamma \over \delta \phi_i(x)}& =& (\Box + \mu^2)\phi_i(x)\nonumber \\
& &+{\lambda\over 6N}[\vec\phi\vec\phi\phi_i(x)+D(x,k;x,k)\phi_i(x)
+2D(x,k;x,i)\phi_k(x)\nonumber\\
         &  & +D(x,k;y,j)V(x^{\prime},k^{\prime};y,j;z,l)
D(x^{\prime},k^{\prime};x,k)D(z,l;x,i)]
\end{eqnarray}
where  
\begin{equation}
D^{-1}(x,a;y,b) = {\delta^2 \Gamma\over \delta\phi_a(x) \delta\phi_b(y)}
\label{dinv}
\end{equation}
and
\begin{equation}
V(x,a;y,b;z,c) = {\delta^3 \Gamma\over \delta \phi_a(x)
\delta\phi_b(y) \delta \phi_c(z)}
\label{vinv}
\end{equation}
It should be noted that $D(x,a;y,b)$ and $V(x,a;y,b;z,c)$ are the
propagator and 3-point vertex respectively only when the above
derivatives are evaluated at the minimum of $\Gamma$ namely, at
$\bar\sigma$ such that
\begin{equation}
{\delta \Gamma\over\delta\phi_i}|_{\phi=v}=0
\end{equation}
Hence $D$ and $V$ in Eqs(\ref{dinv}) and (\ref{vinv}) are $\phi$-dependent.\\
It will prove useful to
display explicitly the behaviour of  the 1PI graphs  in the large--N limit. 
For example, the four-point vertex
has a $1/N$ dependence for large N. Likewise, the expectation value
of the order parameter has a $N^{1/2}$ behaviour,and hence the 
3-point functions in the ordered phase  will have a $N^{-1/2}$ factor.

Now consider the SD equations for $\Gamma^2_{\sigma\sigma}$.
This equation is best represented
diagrammatically as in Fig. 2. 
\begin{figure}[htb]
\begin{center}
\mbox{\epsfig{file=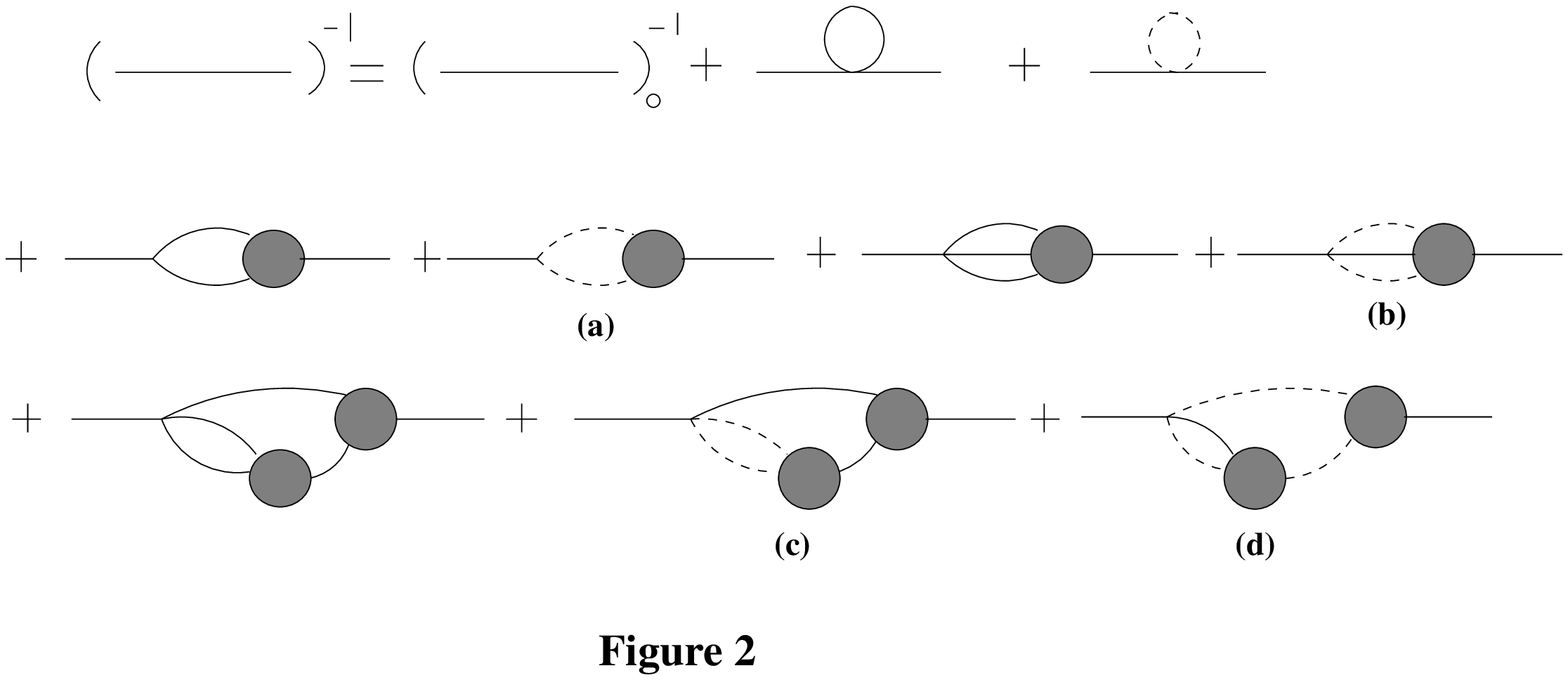,width=8truecm,height=6truecm,angle=-90}}
\caption{ Schwinger-Dyson equations for the $\sigma$ two-point
function }
\label{Fig 2.}
\end{center}
\end{figure}
In these figures solid lines
represent exact $\sigma$-propagators and dashed lines exact
$\pi$-propagators( except the lines with subscript `0'  denoting tree
level propagators ), and all blobs
represent exact vertices.\\

We assume some infrared regulator like the external field H. Then
$m_{\pi}$, the Goldstone boson mass is not zero and we can safely study
the zero momentum behaviour of various graphs. The degree of infrared
divergence is reflected by the leading $m_{\pi}$-dependence. 
Our input is that the exact pion propagator has the behaviour
$(k^2-m_{\pi}^2)^{-1}$ for $k \rightarrow 0$, i.e. we put in the 
information that the pion has a canonical behaviour in the infrared
and study the consequences for the $\sigma$-propagator.Diagram
2(a) contributing to $\Gamma^2_{\sigma\sigma}$ would behave like
$m_{\pi}^{d-4}$ if $\Gamma_{\sigma\pi\pi}$ were O(1) in the IR (and
consequently the total $\Gamma^2_{\sigma\sigma}$ would also be O(1) due
to the Ward identity (eqn \ref{A})). This contribution is of order O(1) in the
large-N count. Thus, either the IR-singular behaviour of 2(a) is
cancelled by other diagrams in fig (2) or {\bf $\Gamma_{\sigma\pi\pi}$
must be at least as soft as $m_{\pi}^{4-d}$}. Diagrams with potential
IR-divergences are 2(b),2(c) and 2(d); apart from all of them being of
O(1/N), none of them is as IR-singular as 2(a). Hence cancellation is
ruled out and $\Gamma_{\sigma\pi\pi}$ must behave at least like 
$m_{\pi}^{4-d+\alpha}$ with $\alpha$-positive semidefinite. Consequently
$\Gamma^2_{\sigma\sigma}$ must vanish as either $m_{\pi}^2$ or
$m_{\pi}^{4-d+\alpha}$ whichever is dominanat.\\
While this proves conclusively that $\chi_L$ must diverge, we still need
to restrict $\alpha$ further. It should be noted that
$\Gamma^2_{\sigma\sigma}$ has terms of order $m_{\pi}^0$ and $O(1)$ in
the large-N count. On the other hand $\Gamma^2_{\sigma\sigma}$ must
vanish as some power of $m_{\pi}$. This is only possible if the
$m_{\pi}^0$ terms are cancelled by such terms from the other diagrams.
As all the diagrams except 2(a) are $O(1/N)$, generically this can only happen
from the ultraviolet contributions to 2(a). Within $1/N$ perturbation theory
at least, such cancellations are not possible in which case the
$m_{\pi}^0$ terms must be generated in the infrared by 2(a) and this is
possible only if $\alpha=0$. This completes the proof based on
SD-equations.\\
\section{Some Universal Features of Multipion Processes}

We have thus established that the divergence of the longitudinal
susceptibility is very general depending only on the goldstone phenomena. In
fact, we have shown on very general grounds that the leading infrared
behaviour of the longitudinal two-point function is completely universal. 

Now we show that the same considerations lead to certain universal features
of multipion scattering amplitudes. Such universal features were alluded to
in Weinberg's article \cite{sw}. Consider, for example, the expression
(67) for the $\pi-\pi$ scattering amplitude at zero momentum. Eqn (74) can
be recast as
\begin{eqnarray}
f^\prime &\sim & H^{\frac{4-d}{2}} \ \ 2<d<4 \\
&&\sim (ln H)^{-1} \ \ for \ \ d=4
\end{eqnarray}

Hence $4v^3f^\prime v^2) >> H$ as $H\rightarrow 0$ and one has
\begin{equation}
A_{4\pi}=\frac{4i}{v^3} (H+\frac{H^2}{4v^3f^"(v^2)}+\cdots)(\delta_{ij}
\delta_{kl}+\cdots)
\label{new}
\end{equation}

On recalling that in the absence of the infrared cut-off H, the scale that
replaces $H$ is $E^2$, we recognise that the leading universal infrared
behaviour $A_{4\pi}\sim E^2$ is indeed the well known result for soft pion
scattering (usually derived in $d=4$). This universal behaviour does not
depend on the space time dimensionality.

However (\ref{new})demonstrates that even the next to leading term is
universal. The above considerations can be applied to multi-pion scatering
amplitudes and one again finds the same universality of the first two terms.
It is also clear from this paper that these results are non-perturbative
in nature. 

The presence of these non-analytic (in energy-momentum variables) terms in
$d=4$ had been noted by Li and Pagels \cite{lp}, the coefficients of which 
in the case
of $\pi-\pi$ scattering had been calculated by Lehmann, and Lehmann and
Trute \cite{lt} using unitarity. Weinberg showed that the application of
renormalisation group can be used to show the universality of these
logarithmic terms. He implicitly uses chiral symmetry. Though superficially
it appears that the universality of the logarithmic terms is a one-loop
result, careful reexamination of Weinberg's demonstartion shows that it is
an all-loop result.

In our approach, the origin of the universality of these subdominant
logarithmic terms is directly related to
universal divergence of the longitudinal susceptibility.

Our large-N derivation, which links the divergence of the $\chi_L$ to the
special properties of the auxiliary field also demonstrates the universal
features in a very straightforward manner.

The multipion scattering amplitudes in this case can be classified into three
categories: a) those involving only $\lambda$ and $\pi$ propagators,
and the $\lambda \pi^2$ vertices. For the scattering amplitude involving
2n pions (including initial and final states ),one needs $n-1 ~\lambda$
and $n-2~\pi$ propagators. Hence the IR-behaviour of all multipion
amplitudes of this class is $\sim p^2$. b) Amplitudes involving one
$n\lambda$
vertex and n $\lambda \pi^2$ vertices. The $n\lambda$-vertex scales like
${p^2}^{d/2-n}$. Hence this contribution to the amplitude scales like
${(p^2)}^{d/2}$, which is less dominant than the behaviour a). Finally, there
is the class c) which consists of amplitudes involving combinations of
multi-$\lambda$ vertices with $m < n$.Their contribution can be seen to
be even less dominant than b).

Because of the $k^2I(k^2)$ factors in the denominators of the
$\lambda$-propagator, one gets a universal correction factor $[1+
\frac{k^2I(k^2)}{v^2}]$ and the leading IR behaviour of these correction
factors is the same in the linear and non-linear $\sigma$-models. This is
precisely the same quantity that governs the universal  divergence of
$\chi_L$. 

We now make a few geenral remarks about conventional chiral perturbation
theory (ChPT) and compare it with our approach.

As we have already stated many times earlier, the leading IR behaviour of
systems with Goldstone bosons is completely universal, and in the sector of
Goldstone bosons only, is represented by the Lagrangian
\begin{equation}
{\cal L}=\frac{1}{2}(\partial_\mu{\vec\pi})^2+\frac{1}{2f_\pi^2}
\frac{({\vec\pi}\times\partial_\mu{\vec\pi})^2}
{(1-\frac{{\vec\pi}^2}{f_\pi^2})^2}
\label{new-1}
\end{equation}
This, following Weinberg, can be cast in the form
\begin{equation}
{\cal L_0}=\frac{1}{2}(D_\mu{\vec\pi})^2
\label{new-2}
\end{equation}

The validity of (\ref{new-1}) is, however, strictly in the extreme infrared.
One of the procedures adopted to extend (\ref{new-1} to regions beyond the
extreme IR is the so-called chiral perturbation theory. In the form that is
mostly practised, ChPT amounts to starting, for example, from a Lagrangian of 
the form 
\begin{equation}
{\cal L}={\cal L_0}+\frac{1}{2}(D_\mu\pi)^2+\frac{g_4^{(1)}}{2}
(D_\mu\pi\cdot D^\mu\pi)^2 +\frac{g_4^{(2)}}{2}(D_\mu\pi\cdot D_\nu\pi)
(D_\mu\pi\cdot D_\nu\pi) + \cdots
\label{new-3}
\end{equation}

Contrary to the attitude of the early days of effective Lagrangians, one now
carries out loop calculations with Lagrangians of the type (\ref{new-2}) and
(\ref{new-3}). Owing to perturbatie non-renormalisability of these
lagrangians, one needs infinitely many counter terms to absorb the various
resulting infinities.. However, to any any desired accuracy in the momenta
involved, one needs only a finite number of counter terms. The structure of
the resulting renormalised Lagrangian whose domain of validity is somewhat
larger that that of (\ref{new-1}) is that it is charaterised by a number of
unknown parameters corresponding to the number of counteer terms used as
well as certain universal parts independent of these unknown coefficients
($f_\pi$ is always treated as known). This is indeed the general structure
argued by Weinberg.

Our analysis that the leading IR behaviour is governed by
the Gaussian fixed point ${\cal G}_{N-1}$ points to a more efficient way of
realising the results of ChPT. This suggestion was implicit in \cite{sw} also.
This consists of perturbing the IR fixed point directly with the help of
irrelevant operators. Their influence on all correlation functions can be
directly computed by the use of the RG techniques. The ultraviolet degrees
of freedom have no role to play in this scheme. This method will be applied
to practical calculations of ChPT elsewhere.
	
\section{Conclusions}
In this paper we have emphasised that the quantum 
fluctuations of the Goldstone bosons have some
drastic effects which are not encountered either in a tree
level or loopwise analysis of the linear or the non-linear
$\sigma$- models.Nevertheless,the leading infrared behaviour of the 
correlation functions again become identical in the linear and
the non-linear versions.It is remarkable that the
longitudinal fluctuations also become `massless' as a
consequence of the masslessness of the Goldstone bosons.
We have exhibited the generality of this phenomenon 
giving  three separate arguments based on i). Renormalization 
group flows, ii). Ward identities, iii). Schwinger-Dyson
equations.  

Heuristic derivations of the soft - pion theorems within the linear
$\sigma$-model are based on the fact that $\sigma$-mass ($\chi_L^{-1}$)
being non--zero, for all processes involving momenta much smaller than this
mass the $\sigma$-propagator can be replaced by $m_{\sigma}^{-2}$. One
may wonder whether the results of this paper would vitiate this and
consequently the soft-pion theorems too. But as demonstrated in the
sections on the large-N analysis as well as the one on Ward identities, 
the soft pion theorems are still valid. The deeper
reason for this is, of course, the renormalisation group point of view
according to which the leading IR-behaviour is controlled by the
Gaussian fixed point.\\

As stressed sufficiently in this article, both the
qualitative and quantitative aspects of the result,as far as the 
susceptibilities are concerned, were known in the
literature from somewhat differing perspectives which were not general
enough. What we have done here is to demonstrate the full generality
of this result as well as extend the analysis to include the 
nonperturbative IR behaviour of correlation functions also. Despite the fact 
that the result concerning $\chi_L$ had appeared in the
literature in different guises , its significance does not appear to have
been well appreciated. \\

We emphasise that in the chiral limit 
the `sigma particle' has a singular 
infrared behaviour (i.e. logarithmic divergence) even
in four dimensions.

Independently of the space-time dimensions, the Goldstone
bosons have canonical dimensions and decouple from each
other and their logitudinal partners in the infrared.We 
pointed out that this can be understood from the Goldstone 
theorem, current algebra and PCAC on the one hand and the
flow towards $O(N-1)$ Gaussian fixed point on the other hand.
This latter point of view is very valuable because one can
easily study the non-leading corrections to the infrared 
behaviour by standard renormalization group methods \cite{z}.
All effects of the microscopic theory could be parametrised
in a few irrelevant and marginal operators around the
$O(N-1)$ Gaussian fixed point and their effects computed.
This approach provides a technically more efficient and 
conceptually simpler alternative to chiral perturbation
theory, which focusses on the infrared divergence of the 
Goldstone bosons instead of the non-renormalizable 
ultraviolet divergences.

We showed that the leading $N$ approximation takes into 
account all the infrared behaviours accuratly, including
that of the longitudinal fluctuations, in contrast to
perturbation theory based on the linear or the non-linear
$\sigma$- models.We developed a technique which allows
explicit computation of the correlation functions even far
away from the infrared region.This provides a useful
non-perturbative technique for obtaining a qualitative 
understanding of the effects of fluctuations.

A convenient tool for understanding longitudinal susceptibility is the
equation of state \cite{z}.In terms of the reduced variables
$y=H/M^{\delta}, x = t/M^{1/\beta}$,where M is the magnetisation, t the
deviation from the critical temperature,and $\delta,\beta$ the standard
exponents, the equation of state takes the universal form\\
\begin{equation}
y = f(x)
\end{equation}
As H tends to zero, $x$ must tend to one of the roots of $f(x)$. With
suitable normalisation this root (corresponding to spontaneous
magnetization) can be arranged to be at $ x =
-1$.Now,the question of whether $\chi_L$ diverges or not can be
answered by knowing the manner in which $f(x)$ approaches zero as $x$
approaches $-1$.If this approach is parametrised as\\
\begin{equation}
f(x) = (1 + x)^{1+p}
\end{equation}
where p is a positive semidefinite number, the behaviour of $\chi_L$ as
H tends to zero is given by\\
\begin{equation}
\chi_L \sim H^{-p/(1+p)}
\end{equation}
In the large-N expansion for arbitrary d  (except d=4 where there are
logarthmic modifications),the equation of state is  
\begin{equation}
y = (1 + x)^{\frac{2}{d-2}}
\end{equation}
as can be proven using the euclidean version of 
eqn(\ref {state}) of sec 4(see also \cite{z}).
As the exponent of $(1 + x)$ is greater than 1 one concludes that the 
longitudinal susceptibility diverges for all $T < T_c$. Modulo the
wavefunction renormalisation constant, $\chi_L$ is just $m_\sigma^{-2}$.

There are interesting consequences of this phenomena both for  finite
temperature QCD as well as QCD at zero temperature. In the case of the
former, under the well argued
scenario that this is a second order transition, Wilczek and Rajagopal
\cite{wilc} conclude that the
transition must belong to the universality class of  $ d=3~~~O(4)$
magnet models. Then the general results of this paper become
applicable.\\ 
In numerical simulations of the QCD chiral phase transition at finite
temperature,the quantity that has been used to characterise the nature
of the phase transition is the so-called $\Delta$-cumulant defined by
\cite{lat94}
\begin{equation}
\Delta={\partial ln \bar\psi\psi \over \partial ln m_q}
\end{equation}
where $\bar\psi\psi$ is the chiral condensate and $m_q$ the current-quark
mass.Thus
\begin{eqnarray}
&T> T_c&~~~~~~~~~\bar\psi\psi \sim m_q~~~~~~\Delta=1 \nonumber \\
&T= T_c&~~~~~~~~~\bar\psi\psi \sim m_q^{1/\delta}~~~~~\Delta=1/\delta 
\label{cum1}
\end{eqnarray}
Below $T_c$,but close to it, one should expect, on the 
basis of the considerations of this paper
\begin{equation}
T< T_c~~~~~~~~~\Delta \sim m_q^{1/2} \label{cum2}
\end{equation}
The same considerations apply to the spontaneously broken phase of zero
temperature QCD. Here one should find
\begin{equation}
\Delta \sim m_q ln m_q \label {cum3}
\end{equation}
characterstic of Goldstone phenomena in four dimensions.\\
Of course, the arguments of Wilczek and Rajagopal that maps the problem
to a classical stat mech problem in $d=3$ is valid exactly at $T_c$. But
continuity would demand that at temperatures in the vicinity of $T_c$ too
the effective dimensionality should be close to 3. Thus, the expectations 
based on eqs (\ref{cum1},\ref{cum2} and \ref{cum3} ) are the correct ones.
This should yield the experimentally interesting signal that for QCD at
finite temperatures $T < T_c$, $\chi_L$ should smoothly go over from
eqn (\ref{cum2}) to eqn(\ref{cum3}) as T is lowered.\\
Indeed, wherever there are Goldstone bosons, one should expect to see the
singular behaviour of $\chi_L$. It is therefore somewhat disappointing 
that clear experimental signatures of this effect have not yet been found
even for the classic ferromagnets.The lack of isotropy in most real life
ferromagnets would make the experimental establishment of this effect a
challenging one. An attempt to see this for ferromagnets has recently 
been made by \cite{ferro}. They should have  been looking for a
$\chi_L \sim H^{-1/2} $ behaviour rather than $\chi_L \sim H^{-1/3}$. The
fact that the behaviour of $\chi_L$ is 'super-universal' in the sense that
it depends only on dimension and the existence of Goldstone modes should be
established in as many diverse experimental circumstances as possible. As
Patashinsky and Pokrovsky \cite{ppb} have pointed out, the effect should also be
observable in the superfluid phase of He.Numerical simulations of the $O(N)$-
models in three dimensions, performed in the ordered phase, should also
establish this effect.\\
Kogut {\em et al.} \cite{ko} report a completely different behaviour
and different exponents for the finite temperature chiral transition in
QCD  in comparison to \cite{wilc}. In particular, 
Kogut {\em et al.} do not find the crucial softening of the
longitudinal mode. More specifically,they  argue that PCAC
restricts the equation of state to be of the form\\
\begin{equation}
f(x) = (1+x)
\end{equation}
There is a fallacy in their argument; PCAC implies the above form of
the equation of state if and only if the longitudinal susceptibility is
assumed to be finite. Only then a Taylor expansion around zero
applied magnetic field is permissible.But our results about the
divergence of $\chi_L$ imply that the assumption of a finite $\chi_L$ is
incorrect. Once allowance is made for a diverging $\chi_L$, PCAC no
longer places any restriction on the form of the equation of state. 
Also, Kogut {\em et al.} carry out their analysis in $d=4$ even at the critical
temperature. \\
\newpage
\noindent {\large \bf Acknowledgements}\\
The main inspiration for this work came from the paper by Wilczek and
Rajagopal. We have benefitted from discussions with many of our
colleagues, in particular, V.Soni, G.Baskaran, Deepak Kumar, P.K. Mitter
J.B. Zuber and E. Seiler.

\end{document}